\documentclass[onecollarge,nospthms]{svjour2}

\usepackage{amsmath,amsfonts,amssymb,mathrsfs}
\usepackage{amssymb,mathrsfs,graphicx}
\usepackage{epsfig,subfigure}
\usepackage{ifthen}
\usepackage{colortbl}
\usepackage{ifthen} 

\definecolor{black}{rgb}{0.0, 0.0, 0.0}
\definecolor{red}{rgb}{1.0, 0.5, 0.5}

\provideboolean{shownotes} 
\setboolean{shownotes}{true} 
\newcommand{\margnote}[1]{
\ifthenelse{\boolean{shownotes}}%
{\marginpar{\raggedright\tiny\texttt{#1}}}%
{}%
}
\newcommand{\hole}[1]{
\ifthenelse{\boolean{shownotes}}%
{\begin{center} \fbox{ \rule {.25cm}{0cm} \rule[-.1cm]{0cm}{.4cm}
\parbox{.85\textwidth}{\begin{center} \texttt{#1}\end{center}} \rule
{.25cm}{0cm}}\end{center}} {} }

\newtheorem{theorem}{Theorem}[section]
\newtheorem{lemma}{Lemma}[section]

\newtheorem{remark}{Remark}[section]
\newtheorem{definition}{Definition}[section]

\newcommand{\bbt}{\mathbb T}

\def\charf {\mbox{{\text 1}\kern-.30em {\text l}}}

\journalname{Journal of Statistical Physics}

\begin{document}

\title{Nonlinear stability of phase-locked states for the Kuramoto model
with finite inertia}

\author{Young-Pil Choi \and Chulho Choi
\and Meesoon Ha \and Seung-Yeal Ha}

\institute{Y.-P. Choi \at Department of Mathematical Sciences,
Seoul National University, Seoul 151-747, Korea \\
\email{freelyer@snu.ac.kr} \and C. Choi \at Department of Physics
and Astronomy, Seoul National University, Seoul 151-747, Korea \\
\email{cch@phya.snu.ac.kr} \and M. Ha (Corresponding author) \at
Department of Physics Education, Chosun University, Gwangju 501-759, Korea \\
\email{msha@chosun.ac.kr} \and S.-Y. Ha \at Department of
Mathematical Sciences, Seoul National University, Seoul 151-747,
Korea \\ \email{syha@snu.ac.kr} }

\date{Received: date / Accepted: date}

\maketitle

\begin{abstract}
We discuss the {\it nonlinear stability} of phase-locked states
for globally coupled nonlinear oscillators with finite inertia,
namely the modified Kuramoto model, in the context of the robust
$\ell^{\infty}$-norm. We show that some classes of phase-locked
states are orbitally $\ell^{\infty}$-stable in the sense that its
small perturbation asymptotically leads to only the phase shift of
the phase-locked state from the original one without changing its
fine structures as keeping the same suitable coupling strength
among oscillators and the same natural frequencies. The phase
shift is uniquely determined by the average of initial phases, the
average of initial frequencies, and the strength of inertia. We
numerically confirm the stability of the phase-locked state as
well as its uniqueness and the phase shift, where various initial
conditions are considered. Finally, we argue that some restricted
conditions employed in the mathematical proof are not necessary,
based on numerical simulation results.
\end{abstract}



\section{Introduction}

Since Kuramoto introduced a mathematical model of coupled
nonlinear oscillators~\cite{Ku} by refining the earlier Winfree's
model~\cite{Wi} to be more mathematically tractable, it has become
a minimal model for collective synchronization phenomena, which
are ubiquitous in real systems ranging from physics to biology.
The original Kuramoto model is very simple but exhibits lots of
rich behaviors including a synchronization transition, where all
the oscillators' phases are tuned by the coupling strength against
the diversity of natural frequencies, and eventually reach a
phase-locked state (frequency entrainment) including in-phase
synchronization with exactly the same value (see~\cite{A-B,Da} for
detailed discussion).

Synchronization is a quite interesting nonlinear phenomenon, which
can yield a dynamic phase transition against the coupling strength
among weakly coupled oscillators for a given natural frequency
distribution. Above the coupling strength threshold, oscillators
become partially phase-locked, which is called partial
synchronization. The existence of partially phase-locked solutions
and their stability were studied by Kuramoto~\cite{Ku} and
Crawford~\cite{Cr}, respectively. As an extension of the
aforementioned works, Aeyels and Rogge addressed the stability
issue for networks of a finite number of oscillators~\cite{A-R},
and Mirollo and Strogatz analyzed the linear stability of
phase-locked states for globally coupled Kuramoto
oscillators~\cite{M-S}. Later on, lots of phase models have been
proposed to describe the dynamic behavior of large populations of
nonlinearly coupled oscillators. Furthermore, it has also been
widely discussed that the nature of synchronization transitions
for the continuum version of the original Kuramoto model is
subject to the shape of the natural frequency distribution
function.
While the synchronization transition has been widely discussed
from the physical point of view, the strict shape of phase-locked
states has not drawn much less attention because it concerns far
from the transition. Of course, the transition is one of the
interesting topics to be studied in many senses, such as critical
behaviors and its universality issue. However, it is mostly
restricted to numerical studies, which implies that there are
always some technical difficulties to reach a correct conclusion
without rigorous mathematical guidelines and/or proofs. In that
sense, it is meaningful if we can report what and how can be
mathematically proven in the perspective of relevant aspects
directly and indirectly to major interests for a given model
before proceeding numerical studies near and at the transition.

In this paper, we address the modified Kuramoto model with finite
inertia and take an interest with the details of phase-locked
states in order to handle the shape of phase-locked states as well
as their uniqueness issue for a fixed distribution of natural
frequencies. Consider an ensemble of finite number of Kuramoto
oscillators with finite inertia. Let $\theta_i = \theta_i(t)$ be
the phase of the $i$-th oscillator and $m$ denotes the strength of
uniform inertia acting on Kuramoto oscillators. In this situation,
the dynamics of $\theta_i$ is governed by the initial valule
problem of second-order ODEs:
\begin{equation} \label{main}
m{\ddot \theta}_i + {\dot \theta}_i = \Omega_i + \frac{K}{N}
\sum_{j=1}^{N} \sin(\theta_j - \theta_i), \quad t>0, \quad
i=1,...,N,
\end{equation}
subject to initial phase-frequency configuration:
\begin{equation} \label{Ku-inertia-ini}
(\theta_i, {\dot \theta}_i)(0) = (\theta_{i0},
\omega_{i0}=\dot{\theta}_{i0}).
\end{equation}
Without loss of generality, we assume that the average of natural
frequencies is zero:
\begin{equation} \label{nat}
\Omega_c := \frac{1}{N}\sum_{i=1}^{N} \Omega_i \equiv 0.
\end{equation}
The system \eqref{main} was first introduced by
Ermentrout~\cite{Er} as a phenomenological model to explain the
slow synchronization of certain biological systems, e.g.,
fireflies of the Pteroptyx malaccae, and this model has been used
to describe various dynamical
systems~\cite{A-S2,D-D-T,P-C,W-C-S1,W-C-S2,W-S0,W-St,W-S}. The
whole review and mathematical results for the Kuramoto model can
be found in~\cite{A-B,S-M1,S-M2,S-S-W}. In the earlier study, the
first author and his collaborators~\cite{C-H-Y} have verified that
for some classes of initial configurations and $mK
\geq\frac{1}{4}$, the speed to the phase-locked states is slower
than that of the original Kuramoto model (without the inertial
term), which illustrates the slow relaxation of Kuramoto
oscillators.

The main novelty of this paper is to provide a simple proof for
the nonlinear stability of some classes of phase-locked states in
globally coupled Kuramoto oscillators with finite inertia. Using
the robust $\ell^{\infty}$-norm, we present the proof under the
suitable conditions on initial phase-frequency configurations,
coupling strengths, and finite inertia. It is noted that
phase-locked states are orbitally $\ell^{\infty}$-stable, which
results in the uniqueness of phase-locked states.

This paper is organized after introduction as follows. In Sec. 2,
we briefly recall some basic definitions and mathematical
structures od the system~\eqref{main}. In Sec. 3, we show the
proof of orbital stability of phase-locked states and provide
extensive numerical simulations, which confirm our analytical
results. Finally, Section 4 is devoted to the summary of main
results and the conclusion of the paper.

\section{Frameworks and main results}
\setcounter{equation}{0} \label{proof}

In this section, we discuss two frameworks depending on the
relative size of inertial strength and present main results on the
orbital stability of phase-locked states. We also present several
mathematical structures of the system \eqref{main}-\eqref{nat},
and recall a second-order Gronwall's lemma from \cite{C-H-Y} which
will be used in the proof of our main result.  \newline

Before we begin our technical discussions,
we first recall several definitions
for the phase-locked state and its orbital stability as follows.
\begin{definition}
\emph{(Phase-locked state and orbital stability)}
\begin{enumerate}
\item {Let $\theta(t) := (\theta_1(t), \cdots, \theta_N(t)) \in
\mathbb{T}^N$ be the solution to the system \eqref{main} -
\eqref{nat}. Then $\theta$ is (strongly) phase-locked if and only
if $\theta$ is an equilibrium solution to the system of ODEs
\eqref{main}, i.e.,
\[ \Omega_i + \frac{K}{N} \sum_{i=1}^N \sin(\theta_j - \theta_i) = 0,
\qquad {\dot \theta}_i = 0. \]} \item{Let $\theta$ and ${\tilde
\theta}$ be two phase-locked solutions to the system \eqref{main}
and \eqref{nat}.  Then, $\theta$ and ${\tilde \theta}$ are
congruent and denoted if and only if there exists a constant
$\beta\in R$ such that
\[ \theta - {\tilde \theta}
= \beta {\mathbb{I}}_N,\] where ${\mathbb{I}}_N:=(1, \cdots, 1)\in
\mathbb{Z}^{N}$.} \item{Let $\theta^e$ be a phase-locked solution
to the system \eqref{main} and \eqref{nat}. Then $\theta^e$ is
orbitally stable if and only if for any initial
phase-configuration $\theta_0$ close to $\theta^e$ in a norm
$||\cdot||$, the solution $\theta(t)$ converges to the phase-shift
of $\theta^e$ in a norm $||~\cdot~||$ as $t \to \infty$, i.e.,
\[ \lim_{t \to \infty} || \theta(t)  -(\theta^e + \beta \mathbb{I}_N)|| = 0,
\quad \mbox{for some constant $\beta$}.
\]}
\end{enumerate}
\end{definition}
Below we add some comments on the above definitions.
\begin{remark}
\emph{}
\begin{enumerate}
\item{In general, a phase-locked state is a traveling profile with
a constant phase velocity $\Omega_c$.} \item{The time-dependent
solution $\theta= \theta(t)$ is a (weakly) phase-locked state if
and only if there exist positive constants
 $C_*, C^* \geq 0$ independent of $t$ satisfying
\[ C_* \leq |\theta_i(t)- \theta_j(t)| \leq C^*, \quad t \geq 0.
\]}
\end{enumerate}
\end{remark}
\subsection{Basic estimates}
We rewrite the system~\eqref{main} as a system of first-order ODEs
for $(\theta_i, \omega_i := {\dot \theta}_i)$:
\begin{align}
\begin{aligned} \label{main-1}
\displaystyle {\dot \theta}_i &= \omega_i, \quad t > 0, ~~i=1, \cdots, N, \\
\displaystyle {\dot \omega}_i &= \frac{1}{m} \Big( -\omega_i +
\Omega_i + \frac{K}{N} \sum_{j=1}^{N} \sin(\theta_j - \theta_i)
\Big).
\end{aligned}
\end{align}
We next introduce macro variables (center-of-mass frame) and micro
variables (fluctuations around macro variables):
\[
\displaystyle  \theta_c := \frac{1}{N} \sum_{i=1}^{N} \theta_i,
\quad \omega_c := \frac{1}{N} \sum_{i=1}^{N} \omega_i, \quad  {\hat
\theta}_i := \theta_i - \theta_c, \quad {\hat \omega}_i := \omega_i
- \omega_c,
\]
Then, macro-variables and micro-variables satisfy, respectively,
\begin{equation} \label{macro}
 {\dot \theta}_c = \omega_c, \quad {\dot \omega}_c =
 -\frac{\omega_c}{m},
\end{equation}
and
\begin{align}
\begin{aligned} \label{micro}
{\dot {\hat \theta}}_i &= {\hat \omega}_i,  \quad m {\dot {\hat
\omega}}_i &= -{\hat \omega}_i + {\hat \Omega}_i + \frac{K}{N}
\sum_{j=1}^{N} \sin({\hat \theta}_j - {\hat \theta}_i).
\end{aligned}
\end{align}
By direct calculations,
 \[
\theta_c(t) =  \theta_c(0) + m \omega_c(0) (1- e^{-\frac{t}{m}}),
\quad \omega_c(t) =  e^{-\frac{t}{m}} \omega_c(0).
\]
Thus, macro-variables converge toward some constant states that
are determined by their initial configurations and the magnitude
of inertia:
\begin{equation} \label{macro-lim}
 \lim_{t \to \infty} (\theta_c(t), \omega_c(t)) = (\theta_c(0) + m \omega_c(0), 0).
\end{equation}
For convenience, we recall the following second-order differential
inequality:
\begin{align}
\begin{aligned}\label{DI-1}
& a {\ddot y} + b {\dot y} + c y + d \leq 0, \quad t > 0, \\
& y(0) = y_0, \quad {\dot y}(0) = y_1,
\end{aligned}
\end{align}
where $a > 0, b, c$ and $d$ are constants.

\begin{lemma}
\emph{\cite{C-H-Y}} Let $y = y(t)$ be a nonnegative $C^2$-function
satisfying the differential inequality \eqref{DI-1}. Then we have
following relations:
\begin{eqnarray*}
&&(i)~b^2-4ac > 0; \cr && y(t) \leq \Big(y_0 + \frac{d}{c} \Big)
e^{-\nu_1 t} + a\frac{e^{-\nu_{2} t}-e^{-\nu_{1} t}}
{\sqrt{b^2-4ac}} \Big(y_1 +\nu_{1}y_0 +
\frac{2d}{b-\sqrt{b^2-4ac}}\Big) - \frac{d}{c}. \cr &&(ii)~b^2-4ac
\leq 0; \cr && y(t)\leq e^{-\frac{b}{2a}t}\Big[y_0 +
\frac{4ad}{b^{2}} + \Big( \frac{b}{2a} y_0 + y_1 + \frac{2d}{b}
\Big)t \Big ] - \frac{4ad}{b^{2}},
\end{eqnarray*}
 where decay exponents $\nu_1$ and $\nu_2$ are given as follows.
\[\nu_{1} := \frac{b+ \sqrt{b^2 - 4ac}}{2a}, \quad \nu_{2} := \frac{b- \sqrt{b^2 - 4ac}}{2a}. \]
\end{lemma}

\subsection{Main results}

In order to present main results regarding the stability of
phase-locked states, we first discuss two frameworks, which depend
on the relative magnitude of inertia $m$ to the strength of
coupling $K$ between oscillators.
\newline

Let us recall a $\ell^p$-norm for a finite-dimensional vector
space. For $\theta \in \mathbb{T}^N$, the $\ell^{p}$-norm of
$\theta$ is defined as follows.
\[
 ||\theta||_{\ell^p} :=  \left \{
\begin{array}{ll}
\displaystyle \Big( \sum_{i=1}^{N} |\theta_i|^p \Big)^{\frac{1}{p}}, \quad   &  p \in [1, \infty), \\
\displaystyle \max_{1 \leq i \leq N} |\theta_i|, \quad & p = \infty.
\end{array}
\right.
\]
Then, phase and natural frequency diameters can be denoted as
\[
D(\theta) := \max_{1 \leq i, j \leq N} |\theta_i - \theta_j| \quad \mbox{and}
\quad D(\Omega) := \max_{1 \leq i, j \leq N} |\Omega_i - \Omega_j|.
 \]
It is noted that for $\theta \in \mathbb{T}^N$ with
$\sum_{i=1}^{N} \theta_i = 0$, $||\theta||_{\ell^\infty}$ and
$D(\theta)$ are equivalent in the sense that
\[ ||\theta||_{\ell^\infty} \leq D(\theta) \leq 2 ||\theta||_{\ell^\infty}. \]

The following two frameworks that concern the magnitude of inertia
were first introduced in~\cite{C-H-Y} for the existence of
phase-locked states to the system~\eqref{main}.
\newline

\noindent $\bullet$ {\bf Framework A}: (Small inertia regime)
\begin{enumerate}
\item {The strength of coupling $K$ and the magnitude of inertia
$m$ satisfy
\[ 0 < D(\Omega) < K, \qquad m K < \frac{D^{\infty}}{4 \sin D^{\infty}}, \]
where $D^{\infty} \in \Big(0, \frac{\pi}{2} \Big)$ is the root of the following trigonometric equation:
\[  \sin x = \frac{D(\Omega)}{K}. \]}
\item {An initial configuration of $(\theta_0, \omega_0)$
satisfies
\[ 0 <  \max \Big \{
D(\theta_0), D(\theta_0) + 2 m {\dot D}(\theta(t)) \Big|_{t = 0}
\Big \}   < D^{\infty}, \]}
\end{enumerate}

\noindent $\bullet$ {\bf Framework B}:  (Large inertia regime)
\begin{enumerate}
\item{The strength of coupling $K$ and the magnitude of inertia
$m$ satisfy
\[ 0 < D(\Omega) < \frac{\pi}{8m}, \qquad m K > \frac{\pi}{8}. \]}
\item{An initial configuration of $(\theta_0, \omega_0)$ satisfies
\[ 0 < \max \Big \{ D(\theta_0), D(\theta_0)
+ 2 m {\dot D}(\theta(t)) \Big|_{t = 0} \Big \}
 < 4 m D(\Omega).
\]}
\end{enumerate}
We also note that for $D^{\infty} \in \Big(0,~\frac{\pi}{2}
\Big)$,
\[ \frac{D^{\infty}}{4 \sin D^{\infty}} < \frac{\pi}{8}. \]
By definitions, Framework A and Framework B correspond some
restricted cases of $mK$. Let ${\mathcal P}$ be the collection of
all phase-locked states evolved from various initial
configurations satisfying either Framework A or Framework B, i.e.,
\begin{eqnarray*}
{\mathcal P} &:=& \{ \theta^e = (\theta^e_1, \cdots, \theta^e_N)
\in \bbt^N~:~ \theta^e := \lim_{t \to \infty} \theta(t),
\quad \mbox{where $\theta(t)$ is the solution to the system \eqref{main} -\eqref{nat} } \cr
                      && \hspace{2cm} \mbox{ with initial datum $\theta_0$
                      satisfying either Framework A or Framework B } \}.
\end{eqnarray*}
The set ${\mathcal P}$ is a proper subset of all possible
phase-locked states for the system \eqref{main} and \eqref{nat}
(see Section 3.2), and phase-locked states in ${\mathcal P}$ have
a diameter strictly less than $\frac{\pi}{2}$.   \newline

We now move onto the main results of this paper, the orbital
stability of phase-locked states.
\begin{theorem}
\emph{(Orbital stability)}\label{OrbitalStability}\\
Suppose either Framework A or Framework B hold, and let $\theta^e$
be a given phase-locked state in ${\mathcal P}$. Then $\theta^e$
is orbitally $\ell^{\infty}$-stable in the sense that for any
perturbed initial configuration ${\tilde \theta}_0$ satisfying the
condition (2) in either Framework A or Framework B, the perturbed
solution ${\tilde \theta}(t)$ satisfies
\[   \lim_{t \to \infty} || {\tilde \theta}(t)
- (\theta^e + (\Delta \theta)^{\infty} \mathbb{I}_N) ||_{\ell^{\infty}} = 0, \]
where the phase-shift $(\Delta \theta)^{\infty}$ is explicitly given by
\[ (\Delta \theta)^{\infty} := \theta^e_c - {\tilde \theta}_c(0)  - m {\tilde \omega}_c(0). \]
\end{theorem}

\begin{remark}
\emph{}

\noindent 1. The conditions of Framework A and Framework B are
independent of the system size $N$. Hence, our results can be
lifted to the kinetic regime via the thermodynamic limit.

\noindent 2. Under both Framework A and Framework B, the initial
phase-frequency configuration $(\theta_0, \omega_0)$ evolves
toward the asymptotic phase-locked state $(\theta^{\infty}, 0)$:
 \[ \Omega_i + \frac{K}{N} \sum_{i=1}^N \sin(\theta^{\infty}_j - \theta^{\infty}_i) = 0, \qquad
\sup_{t \geq 0} D(\theta(t)) < \frac{\pi}{2}. \]

\noindent 3. For the case of $m = 0$ (the original Kuramoto
model), the orbital stability of phase-locked states is provided
in~\cite{C-H-J-K} using the $\ell^{1}$-contraction theory:
\[
\lim_{t \to \infty} || \tilde{\theta}(t) - (\theta^{e}
+ (\Delta \theta)^{\infty} \mathbb{I}_N) ||_{\ell^1} = 0.
\]
However, we cannot apply this estimate to the system \eqref{main}
due to inertia. Thus, we employ a new estimate to include the
previous result and also cover the inertial effect, which is based
on $\ell^{\infty}$-metric.

\noindent 4. Theorem 2.1 implies that a phase-locked state with
the phase diameter $D(\theta)$ strictly less than $\frac{\pi}{2}$
is unique up to the phase shift. Let $\theta$ and $\tilde{\theta}$
be the two phase-locked states emerged from initial data
$(\theta_0,\omega_0)$ and $(\tilde{\theta}_0, \tilde{\omega}_0)$,
respectively. Suppose that
\[
D(\theta^{\infty}),~ D(\tilde{\theta}^{\infty}) < \frac{\pi}{2}.
\]
Then, by the same argument as in Theorem 2.1, we have
\[
\theta^{\infty} - \tilde{\theta}^{\infty} = \Big( \theta_c(0) -
\tilde{\theta}_c(0) + m (\omega_c(0) - \tilde{\omega}_c(0)) \Big)
\mathbb{I}_N.
\]
\end{remark}

\section{Nonlinear stability of phase-locked states}
\setcounter{equation}{0}

In this section, we claim the orbital nonlinear stability of the
phase-locked states in the $\ell^{\infty}$ norm as providing some
analytical proof, which is also numerically confirmed.

\subsection{The proof of Theorem 2.1}
Suppose either Framework A or Framework B hold, and let $\theta^e$
be a given phase-locked state in ${\mathcal P}$ with $D(\theta^e)
< \frac{\pi}{2}$ and ${\tilde \theta}_0$ be a perturbed
phase-configuration of $\theta^e$. Then, thanks to Theorem 2.1,
the perturbed configuration ${\tilde \theta} = {\tilde \theta}(t)$
with the initial configuration ${\tilde \theta}_0$ satisfies
\[ D(\theta^e) + D({\tilde \theta}(t)) < \pi, \qquad t \geq 0. \]
We set
\begin{eqnarray*}
\alpha_i &:=& \theta^e_{i} - {\tilde \theta}_i, \quad  \alpha_c(t) := \frac{1}{N} \sum_{i=1}^{N} \alpha_i(t),
\quad \hat{\alpha}_{i}(t) := \alpha_{i}(t) - \alpha_{c}(t), \cr
\hat{\alpha}_{M} &:=& \max_{1 \leq i \leq N} \hat{\alpha}_{i},
\quad \hat{\alpha}_{m} := \min_{1 \leq i \leq N} \hat{\alpha}_{i}, \quad D(\alpha(t))
= \hat{\alpha}_{M} - \hat{\alpha}_{m}.
\end{eqnarray*}
Note that
\begin{align}
\begin{aligned} \label{Est}
\alpha_{c}(t) &= \theta^e_{c} - \tilde{\theta}_{c}(t) =
\theta^e_{c} - \tilde{\theta}_{c}(0) - m(1 - e^{-\frac{t}{m}})
\tilde{\omega}_{c}(0), \\
              &\to \theta^e_{c} - \tilde{\theta}_{c}(0) - m \tilde{\omega}_{c}(0), \quad \mbox{as $t \to \infty$}, \\
|\alpha_{j}(t) - \alpha_{i}(t)| &= |\theta^e_{j} - \tilde{\theta}_{j}(t) - (\theta^e_{i} - \tilde{\theta}_{i}(t))|
\leq D(\theta^e) + D({\tilde{\theta}}(t)) <\pi,
\quad \mbox{for} \quad t \geq 0.
\end{aligned}
\end{align}
By simple calculations, we obtain
\begin{equation}\label{stable-1}
m\frac{d^2 \hat{\alpha}_i}{dt^2} + \frac{d\hat{\alpha}_i}{dt} =
\frac{2K}{N} \sum_{j=1}^{N} \cos \Big( \frac{\theta^e_{j} -
\theta^e_{i}}{2} + \frac{{\tilde \theta}_j -{\tilde \theta}_i}{2}
\Big)
 \sin \Big( \frac{\hat{\alpha}_j - \hat{\alpha}_i}{2} \Big).
\end{equation}
$\bullet$ Step A (Derivation of Gronwall's inequality for
$D(\alpha)$):

It follows from \eqref{stable-1} that
\begin{align}
\begin{aligned} \label{EQ-1}
m\frac{d^2 \hat{\alpha}_{M}}{dt^2}+\frac{d\hat{\alpha}_{M}}{dt} &=
\frac{2K}{N} \sum_{j=1}^{N} \cos \Big( \frac{\theta^e_{j} -
\theta^e_{i}}{2}
+ \frac{{\tilde \theta}_j -{\tilde \theta}_i}{2} \Big)  \sin \Big( \frac{\hat{\alpha}_j - \hat{\alpha}_M}{2} \Big) \\
&\leq \Big( \frac{ K \sin 2 D_0^{av}}{D_0^{av} N} \Big) \sum_{j=1}^{N} \Big( \frac{\hat{\alpha}_j - \hat{\alpha}_M}{2} \Big) \\
&= -\frac{K \sin 2 D_0^{av}}{2 D_0^{av}} {\hat a}_M,
\end{aligned}
\end{align}
where we used the fact $-\pi < \hat{\alpha}_{j} - \hat{\alpha}_{M} \leq 0$ to find
\begin{eqnarray*}
&& D_0^{av} := \frac{D(\theta^e) + D({\tilde \theta}_0)}{2}, \quad
\sum_{i=1}^{N} \hat{\alpha}_{i} = 0, \cr && \sin \Big(
\frac{\hat{\alpha}_j - \hat{\alpha}_M}{2} \Big) \leq \Big(
\frac{\sin D_0^{av}}{D_0^{av}} \Big)
 \Big( \frac{\hat{\alpha}_j - \hat{\alpha}_M}{2} \Big).
\end{eqnarray*}
Similarly, we find
\begin{equation} \label{EQ-2}
m\frac{d^2 \hat{\alpha}_{m}}{dt^2}+\frac{d\hat{\alpha}_{m}}{dt}
\geq - \frac{K \sin 2 D_0^{av}}{2 D_0^{av}} {\hat a}_m.
\end{equation}
We combine the estimates \eqref{EQ-1} and \eqref{EQ-2} to find
\begin{equation} \label{EQ-3}
m\frac{d^2 D({\alpha})}{dt^2}+\frac{d D(\alpha)}{dt} + \bar{K}
D({\alpha}) \leq 0, \qquad \bar{K} := \frac{K \sin 2 D_0^{av}}{2
D_0^{av}}.
\end{equation}
$\bullet$ Step B (Decay estimates of $D(\alpha)$):

We apply Lemma 2.1 for \eqref{EQ-3} to obtain
\[
D(\alpha(t)) \leq \left\{
                    \begin{array}{ll}
                     D({\alpha}_0) e^{-\mu_{1}t} + m\frac{e^{-\mu_{2}t} - e^{-\mu_{1}t}}{\sqrt{1 - 4m\bar{K}}}
  \Big( {\dot D}(\alpha_0) + \mu_{1} D({\alpha}_0) \Big), &1 - 4m\bar{K} > 0,     \\
                      e^{-\frac{t}{2m}}\Big[  D({\alpha}_0)
                      + \Big(\frac{1}{2m} D({\alpha}_0) + \dot{D}({\alpha}_0) \Big ) t \Big],
 & 1 - 4m\bar{K} \leq 0,
                    \end{array}
                  \right.
\]
where
\[
\mu_{1} = \frac{1 + \sqrt{1 - 4m \bar{K}}}{2m}, \quad \mu_{2} =
\frac{1 - \sqrt{1 - 4m\bar{K}}}{2m}.
\]
Hence, for any $\varepsilon \in \Big(0, \frac{1}{2m} \Big)$, we
have
\[
D({\alpha}(t)) \approx {\mathcal O}(1) e^{-\lambda(\varepsilon) t}
\quad \mbox{for~large~time} t, \quad \lambda(\varepsilon)
 := \min \Big \{ \mu_2, \frac{1}{2m}- \varepsilon \Big \}. \]
We set
\[ (\Delta \theta)^{\infty} := \theta^e_c - {\tilde \theta}_c(0)  - m{\tilde \omega}_c(0). \]
Using the triangle inequality, we get the following results:
\begin{eqnarray*}
| |\theta^e - {\tilde \theta}(t) -(\Delta \theta)^{\infty} \mathbb{I}_N||_{\ell^{\infty}}
 &=& ||\alpha(t) -(\Delta \theta)^{\infty} \mathbb{I}_N||_{\ell^{\infty}} \cr
&\leq& ||\alpha(t) - \alpha_c(t)| + |\alpha_c(t) -(\Delta
\theta)^{\infty} \mathbb{I}_N||_{\ell^{\infty}} \cr
                                 &\leq& 2D(\alpha(t)) + ||\alpha_c(t) -(\Delta \theta)^{\infty}
                                 \mathbb{I}_N||_{\ell^{\infty}},
\end{eqnarray*}
which implies
 \[ \lim_{t \to \infty} || {\tilde \theta}(t) -
 ( \theta^e +  (\Delta \theta)^{\infty} \mathbb{I}_N)||_{\ell^{\infty}} = 0. \]
This completes the proof.

\subsection{Numerical simulations}

\begin{figure}[]
\centering
\includegraphics[angle=-90, width=0.9\linewidth]{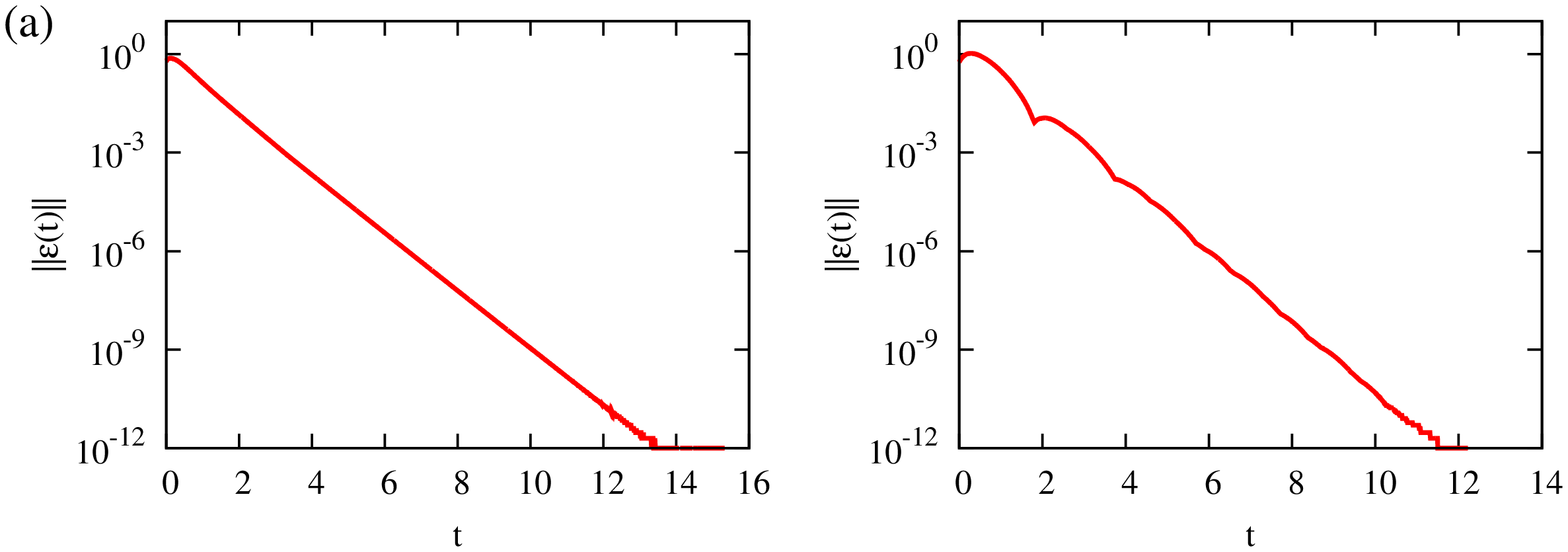}
\includegraphics[angle=-90, width=0.9\linewidth]{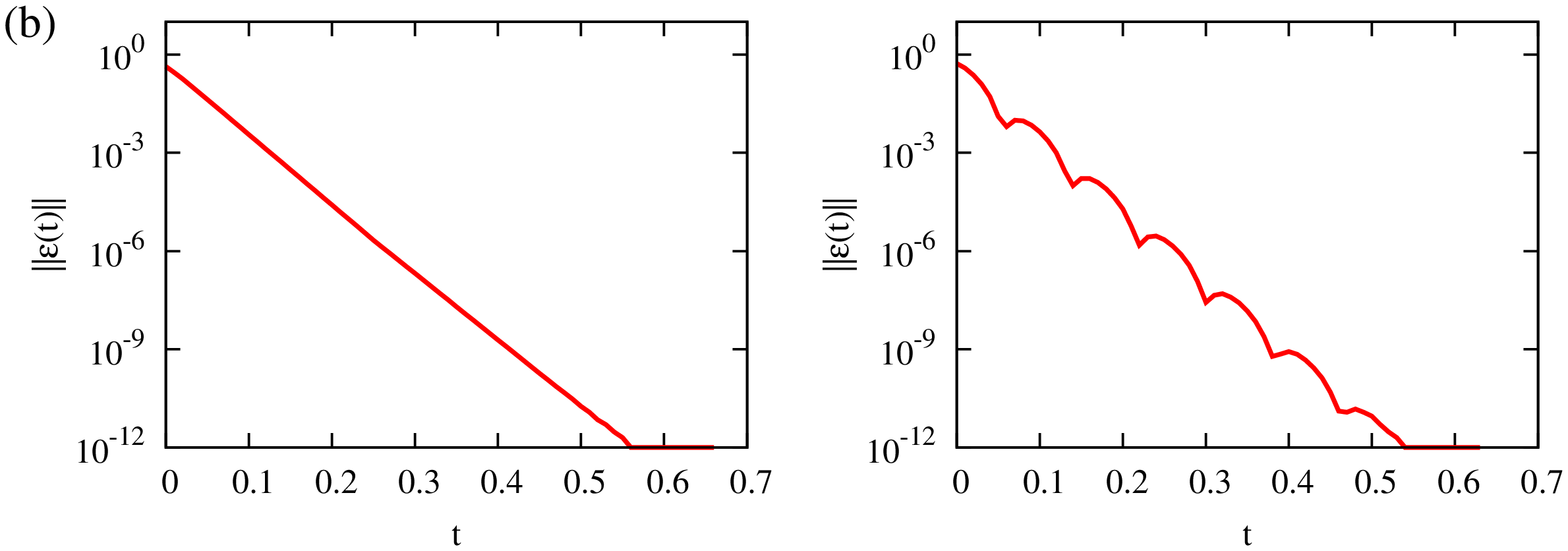}
\caption{   \label{fig:frameAB_E_vs_t}
    (Color online) Semi-logarithmic plots of
    the relative error, $||{\mathcal E}(t)||$, against time, $t$,
    which show the exponential decay for various conditions:
    $N=100$ and 100 realizations from
    $\{\theta_i(0)\}\subset [-\frac{\pi}{10},\frac{\pi}{10}]$, and
    $\{\omega_i(0)\}\subset [-\pi, \pi]$. We test two different distributions of natural
    frequencies, (a) for the uniform distribution, Framework A (left) with ($m$=0.1, $K$=2.0)
    and Framework B (right) with ($m$=0.2, $K$=2.0) and (b) for Cauchy (Lorentzian) distribution,
    Framework A (left) with ($m$=0.004, $K$=40.0)
    and Framework B (right) with ($m$=0.01, $K$=40.0). }
\end{figure}
\begin{figure}[]
\centering
\includegraphics[angle=-90, width=0.9\linewidth]{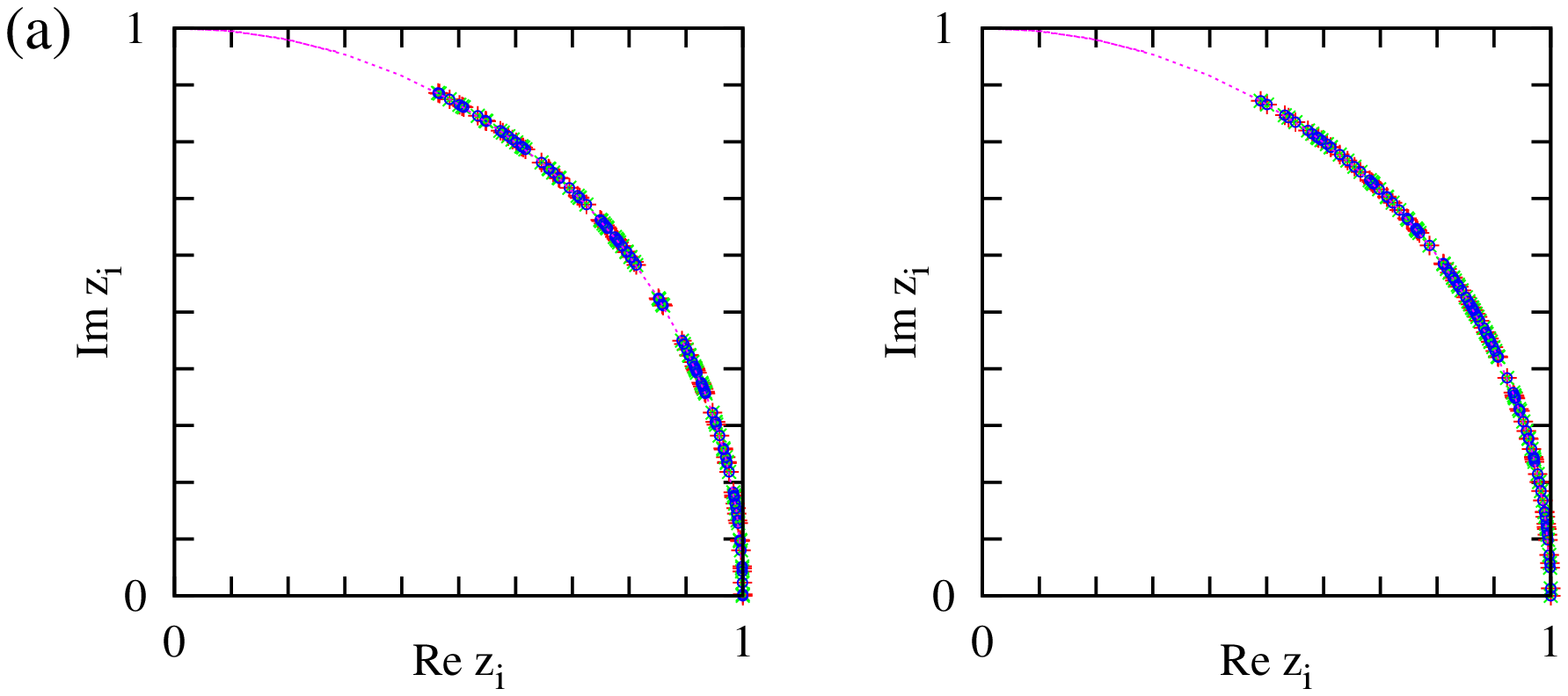}
\includegraphics[angle=-90, width=0.9\linewidth]{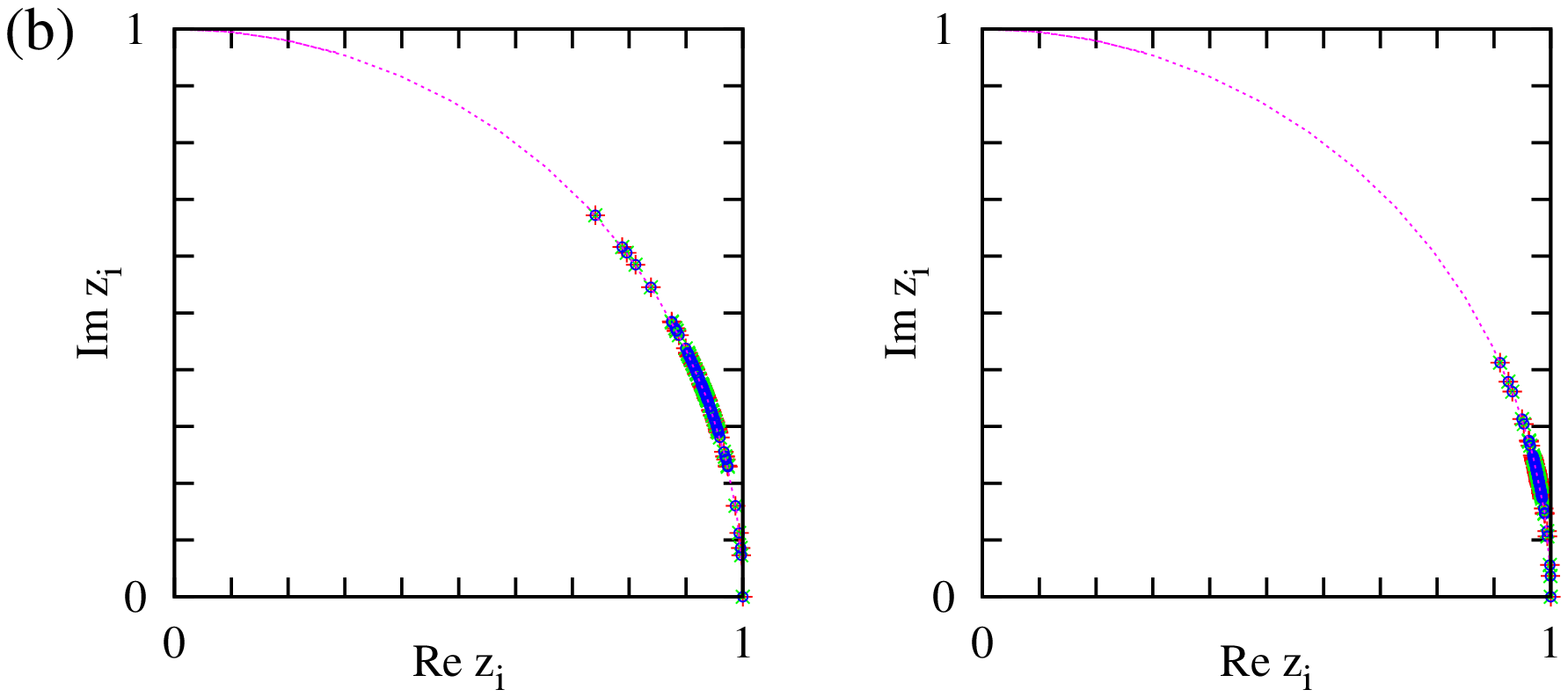}
\caption{   \label{fig:frameAB_last_phase}
    (Color online) Three different phase-locked states ($\circ,~\times,~+$) on complex plane,
    which coincide with one another by the rotation with the relative
    phase-shift for $N=100$, $\{\theta_i(0)\}\subset [-\frac{\pi}{10},\frac{\pi}{10}]$, and
    $\{\omega_i(0)\}\subset [-\pi, \pi]$. Two different distributions of natural frequencies are
    tested with the same setups of Fig.~\ref{fig:frameAB_E_vs_t}, summarized in Table~\ref{table:All-Conditions}.
    The dashed line on the complex plane is guided for eyes as drawing an unit circle
    whose center is located at the origin point.}
\end{figure}
\begin{figure}[]
\centering
\includegraphics[angle=-90, width=0.9\linewidth]{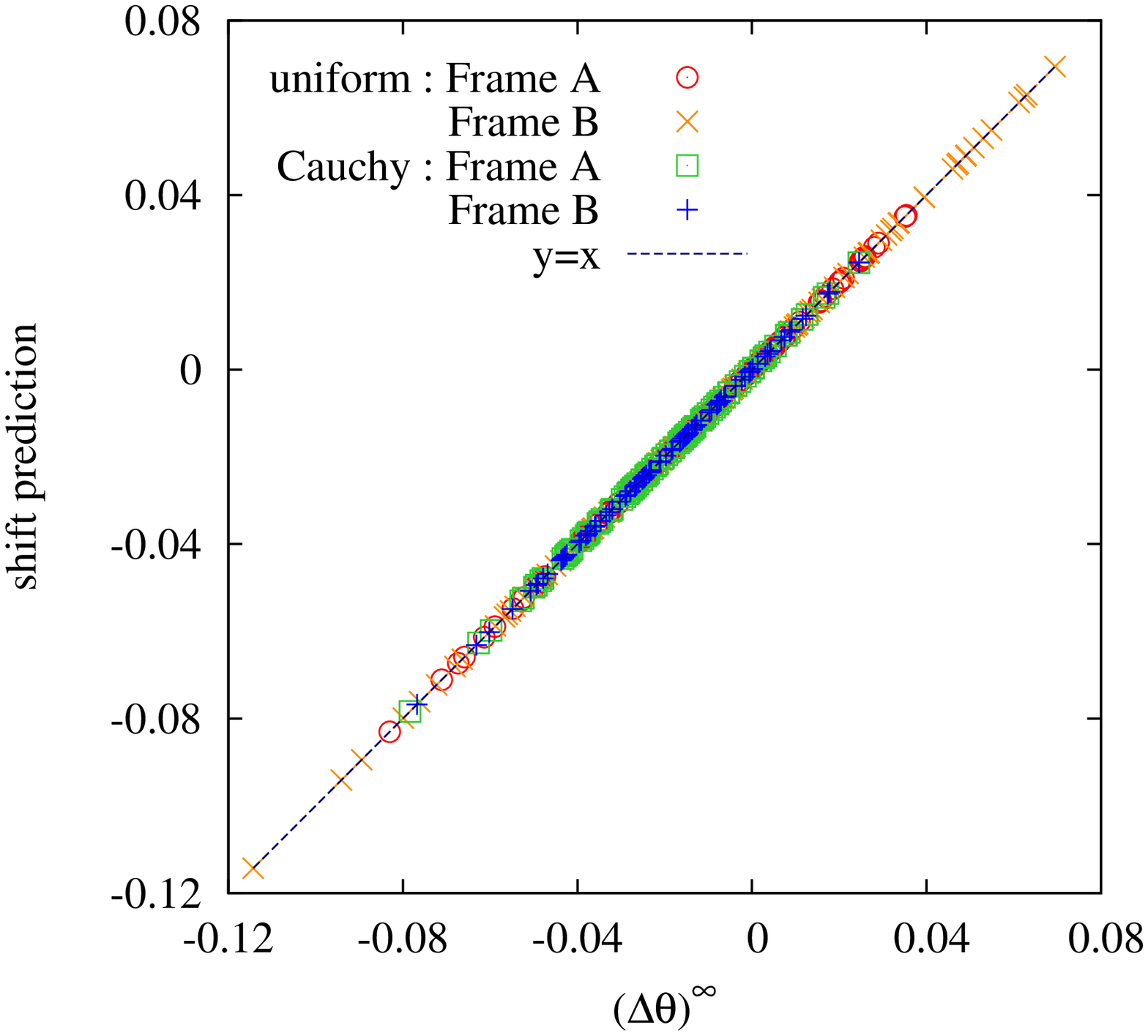}
\caption{   \label{fig:frameAB_shift}
    (Color online) For phase-locked states, the analytic estimate (Theorem~\ref{OrbitalStability})
    of the relative phase-shift is confirmed in Framework A and Framework B
    with two distributions of natural frequencies: the uniform distribution and the Cauchy distribution,
    respectively. We plot all the cases of Figs.~\ref{fig:frameAB_E_vs_t} and ~\ref{fig:frameAB_last_phase},
    summarized in Table~\ref{table:All-Conditions}. Our numerical simulations and analytical estimates
    are in perfect agreement with the guide line of $y=x$.}
\end{figure}
\begin{table}
\centering \caption{All the conditions and parameter sets used in
our simulation tests for random 100 realizations of
$\theta(0)\in[-\pi/10,\pi/10]$ and $\omega(0)\in[-\pi,\pi]$:}
\label{table:All-Conditions}
\begin{tabular}{l|*{4}{c}}  \hline\hline
  & $\Omega$ Distribution & Framework & m & K  \\  \hline
Fig1,2(a)left   & uniform($[-1,1]$) & A & 0.1   & 2.0    \\
Fig1,2(a)right  & uniform($[-1,1]$) & B & 0.2   & 2.0     \\
Fig1,2(b)left   & Cauchy($\gamma=1$)    & A & 0.004 & 40.0   \\
Fig1,2(b)right  & Cauchy($\gamma=1$)    & B & 0.01  & 40.0    \\
Fig4,5(a)left   & uniform($[-1,1]$) & - & 1.0   & 3.0     \\
Fig4,5(a)right  & uniform($[-1,1]$) & - & 0.2   & 2.0     \\
Fig4,5(b)left   & Cauchy($\gamma=1$)    & - & 0.5   & 40.0    \\
Fig4,5(b)right  & Cauchy($\gamma=1$)    & - & 0.1   & 20.0    \\
\hline\hline
\end{tabular}
\end{table}
\begin{table}
\centering
\caption{The values in the center of mass frame for
selected three initial conditions among 100 realizations for
Figs.~\ref{fig:frameAB_last_phase} and
~\ref{fig:Noframe_last_phase}.} \label{table:Three-Cases}
\begin{tabular}{c|cc}
\hline\hline
Sample No.  & $\theta_c(0)$ & $\omega_c(0)$\\
\hline
1   & 0.020411279668    & -0.056797709228   \\
2   & 0.015377250725    & 0.283280670045    \\
3   & 0.022909592396    & -0.004959481400   \\
\hline\hline
\end{tabular}
\end{table}
\begin{figure}
\centering
\includegraphics[angle=-90, width=0.9\linewidth]{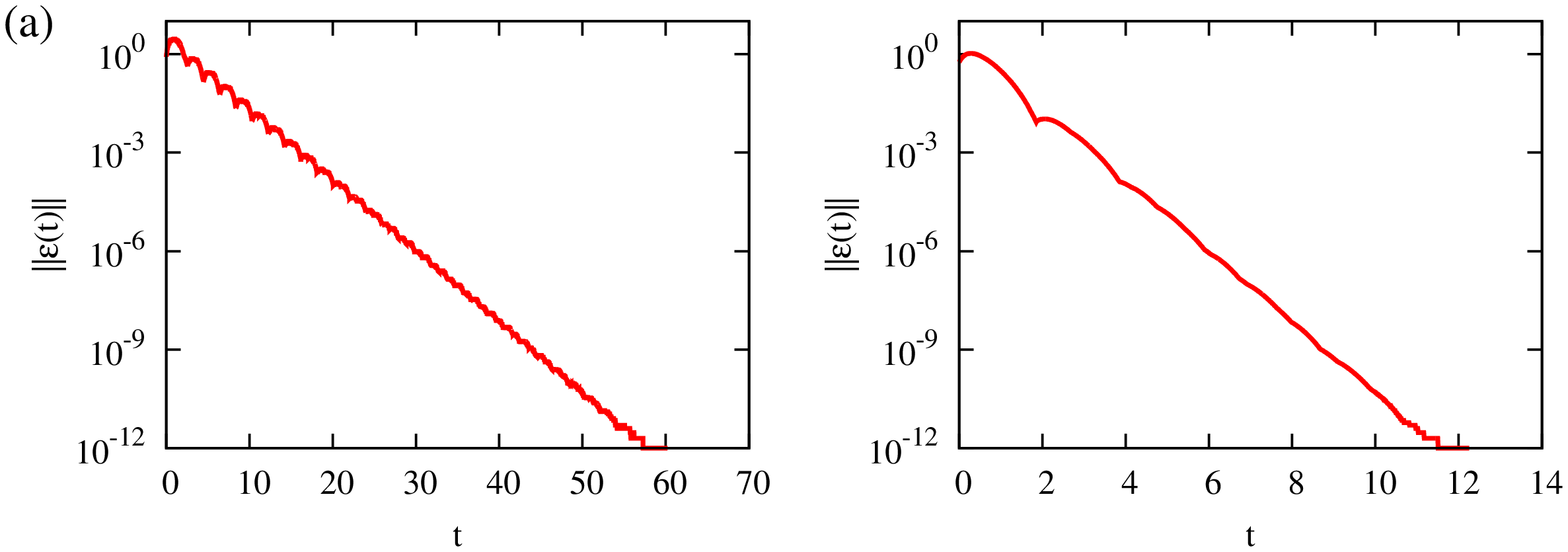}
\includegraphics[angle=-90, width=0.9\linewidth]{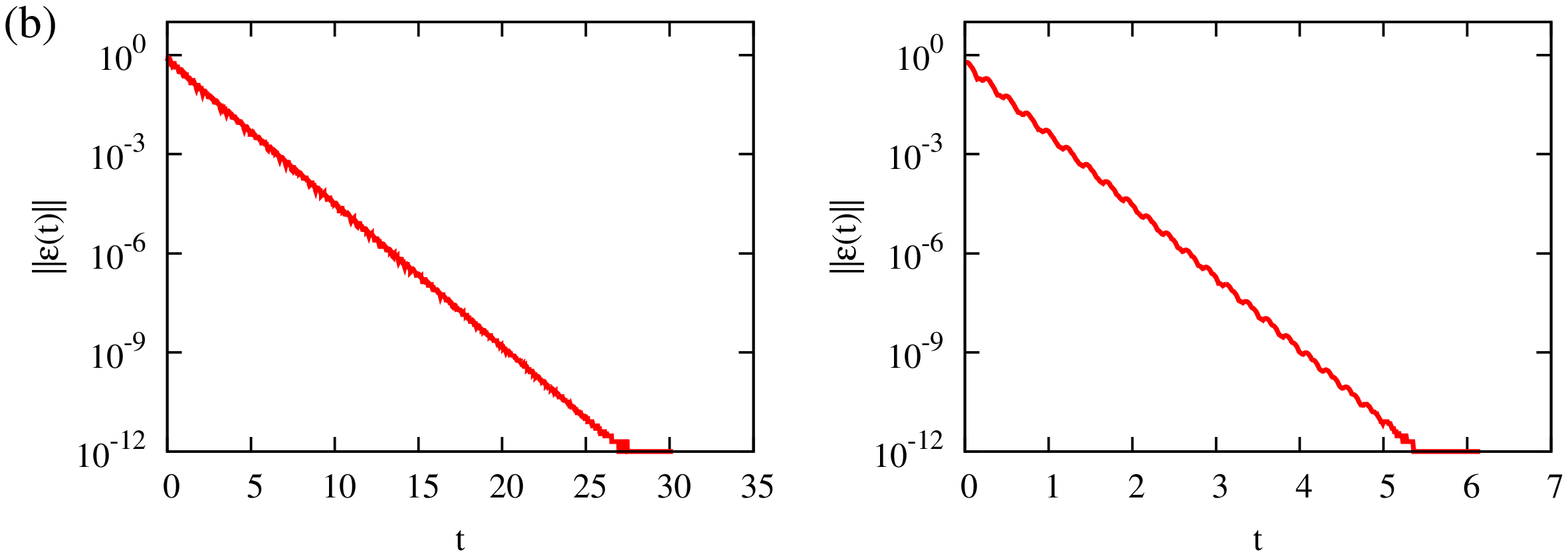}
\caption{   \label{fig:Noframe_E_vs_t}
    (Color online) Semi-logarithmic plots of the relative error, $||{\mathcal E}(t)||$, against time, $t$,
    which show the exponential decay for various conditions: $N=100$ and 100 realizations from
    $\{\theta_i(0)\}\subset [-\frac{\pi}{10},\frac{\pi}{10}]$, and  $\{\omega_i(0)\}\subset [-\pi, \pi]$.
    We test two different distributions of natural frequencies, (a) for the uniform
    distribution, ($m$=1.0, $K$=3.0) (left) and ($m$=0.2,
    $K$=2.0) (right), and (b) for the Cauchy distribution, ($m$=0.5, $K$=40.0)
    (left) and ($m$=0.1, $K$=20.0) (right). Note that these
    parameter setups satisfy neither Framework A nor Framework B.}
\end{figure}
\begin{figure}
\centering
\includegraphics[angle=-90, width=0.9\linewidth]{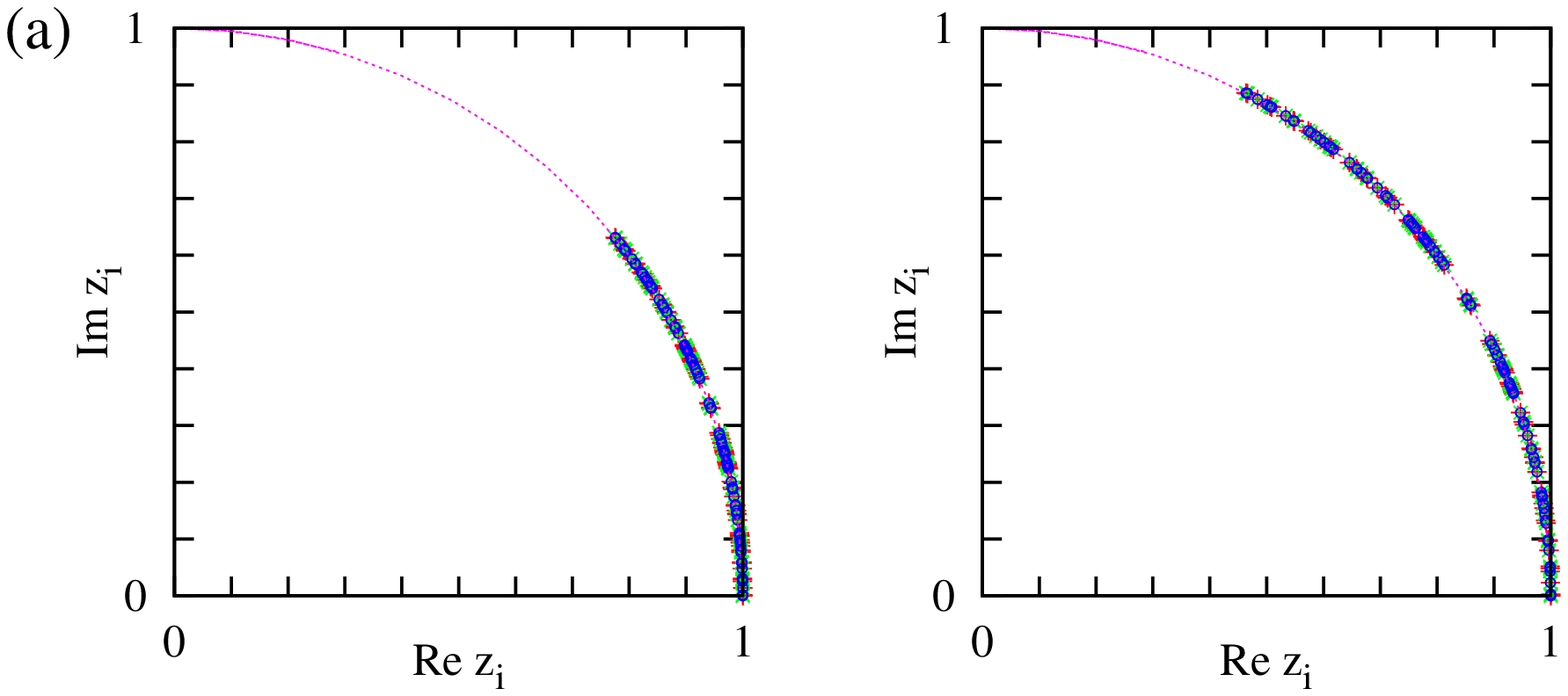}
\includegraphics[angle=-90, width=0.9\linewidth]{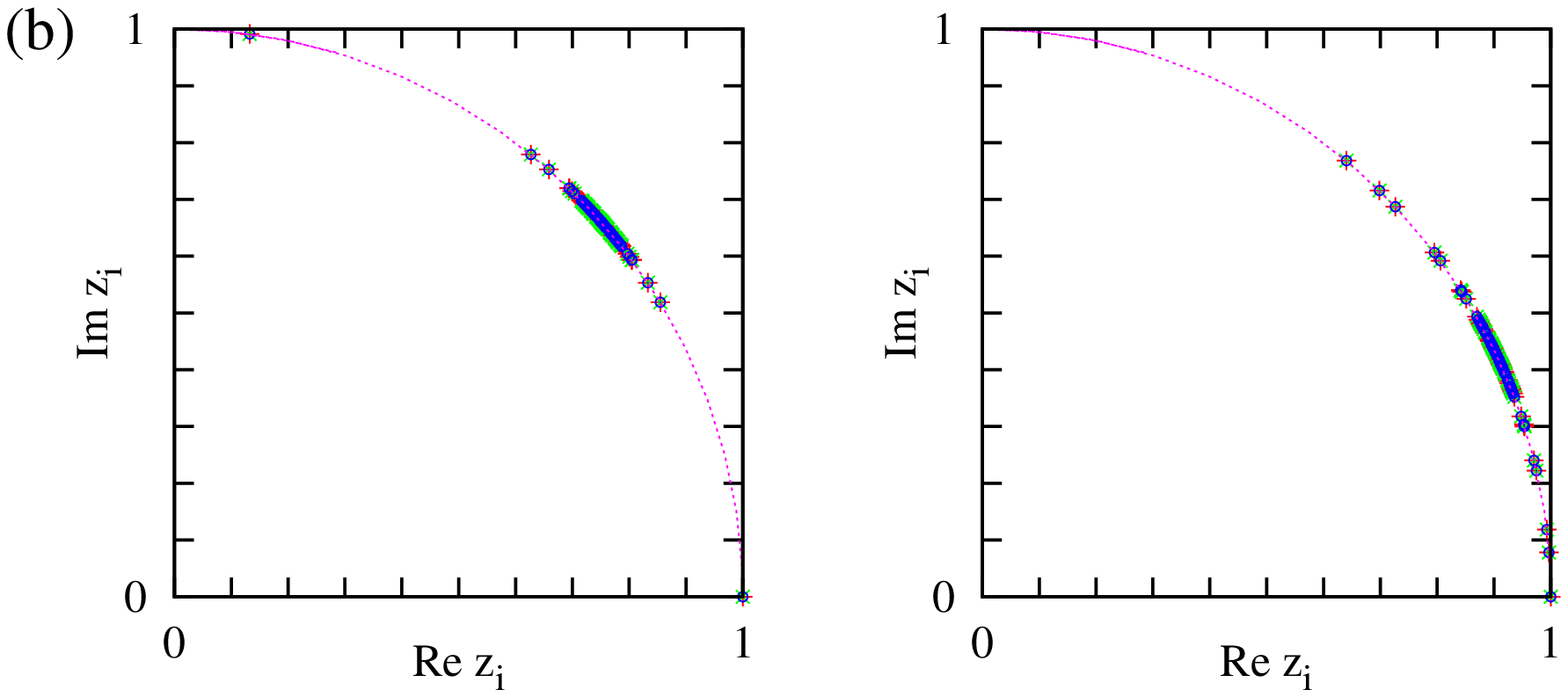}
\centering
\caption{   \label{fig:Noframe_last_phase}
    (Color online) Three different phase-locked states ($\circ,~\times,~+$) on complex plane,
    which coincide with one another by the rotation with the relative
    phase-shift for $N=100$, $\{\theta_i(0)\}\subset [-\frac{\pi}{10},\frac{\pi}{10}]$, and
    $\{\omega_i(0)\}\subset [-\pi, \pi]$. Two different distributions of natural frequencies are
    tested with the same setups of Fig.~\ref{fig:Noframe_E_vs_t}, summarized in Table~\ref{table:All-Conditions}.
    The dashed line on the complex plane is guided for eyes
    as drawing an unit circle whose center is located at the origin point.}
\end{figure}
\begin{figure}
\centering
\includegraphics[angle=-90, width=0.9\linewidth]{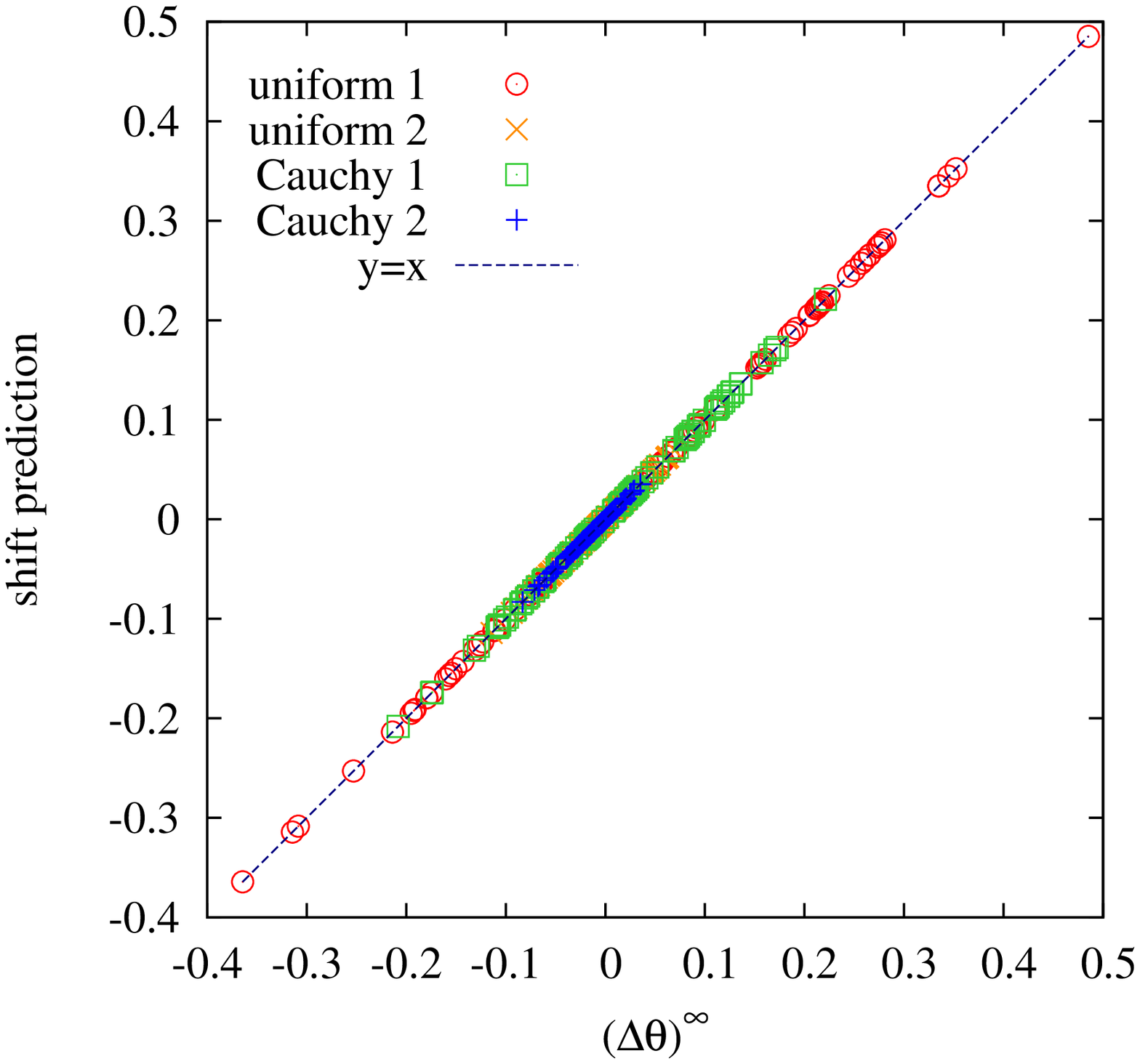}
\centering
\caption{   \label{fig:Noframe_shift}
    (Color online) For phase-locked states, the analytic estimate (Theorem~\ref{OrbitalStability})
    of the relative phase-shift is confirmed in Framework A and Framework B
    with two distributions of natural frequencies: the uniform distribution and the Cauchy distribution,
    respectively. We plot all the cases of Figs.~\ref{fig:Noframe_E_vs_t} and ~\ref{fig:Noframe_last_phase},
    summarized in Table~\ref{table:All-Conditions}. Our numerical simulations and analytical estimates
    are in perfect agreement with the guide line of $y=x$.}
\end{figure}

In order to confirm our analytic proof and its validity, we
numerically test various situations, which is delineated as
follows. We employ the 4th order Runge-Kutta method to numerically
integrate the modified Kuramoto model, Eq.~(\ref{main}), where
\[  \mbox{the time step}~~ \Delta t=10^{-2} \quad \mbox{and} \quad N = 100. \]
As the natural frequency distribution, we consider
\[   g_{c}(\Omega)= \frac{1}{ \pi (\Omega^2+1)}~:~\mbox{Cauchy distribution},
\qquad g_u(\Omega) = \frac{1}{2} {\bf 1}_{[-1, 1]}~:~\mbox{uniform
distribution}. \] One Half of natural frequencies are randomly
chosen from the given analytic distribution form in the positive
part $\Omega> 0$, and the other half is set to be negative of
chosen values, so that the mean of $\Omega_i$ set to be 0. Initial
configurations of $\{\{\theta_i(0)\},
\{\omega_i(0)=\dot\theta_i(0)\}\}$ are uniformly chosen from the
intervals: $\theta_i\in\Big[-\frac{\pi}{10}, \frac{\pi}{10} \Big]$
and $\omega_i\in[-\pi,\pi]$ at time $t=0$, respectively.

The following procedure is how we obtain numerical results in this
paper: Prepare several sets of natural frequencies and initial
conditions, $\{\{\Omega_i\}, \{\theta_i(0)\}, \{\omega_i(0)\}\}$.
Among them, pick up a set, $\{\{\Omega^{\rm locked}_i\},
\{\theta^{(0)}_i(0)\}, \{\omega^{(0)}_i(0)\}\}$ with which
oscillators converge to a phase-locked state. In this paper, we
test various initial setups, including the simplest ones: the case
for initially static, $\omega_i(0)=0$ for any oscillators, and the
case for Kuramoto-type dynamic, $\omega_i(0) = \Omega_i +
\frac{K}{N}\sum_{j=1}^{N}{\sin(\theta_j-\theta_i)}$. We find that
for random initial configurations, there are no significant
difference between numerical results in the steady state. As a
result, we can detect phase-locked states as checking the
condition $|\theta_i(t+dt)-\theta_i(t)|<\varepsilon$ at every time
step, where we set $\varepsilon=10^{-13}$. If all phases satisfy
the given condition, we assume that the system reaches the
neighborhood of phase-locked states.

In Fig.~\ref{fig:frameAB_E_vs_t}, we first check out the 4th
statement in Remark 2.2 with Framework A and Framework B as stated
in Sec. 3, where without loss of generality, we take
$\theta(0)\in\Big[-\frac{\pi}{10},\frac{\pi}{10}\Big]$ and
$\omega(0)\in [-\pi,\pi]$ and measure the relative error from one
phase-locked state to one another as follows:
\[ {\mathcal E}(t) : =  {\mathcal E}(t) - {\mathcal E}_c(0),
\quad {\mathcal E}(t) : = \theta(t) - \tilde{\theta}(t), \quad
{\mathcal E}_c(0) := \Big(\theta_c(0) - \tilde{\theta}_c(0) + m
(\omega_c(0) - \tilde{\omega}_c(0))\Big) \mathbb{I}_N, \] where
$\tilde{X}$ and $X_c$ correspond to the perturbed value from one
selected setting and the value in the ``center of mass" frame,
respectively. The values used in our calculation are shown in
Table~\ref{table:Three-Cases}. As expected, the relative error
$||{\mathcal E}(t)||$ decays exponentially fast to zero, which is
the numerical verification of Theorem~\ref{OrbitalStability} as
well.

Once $\{\Omega_i\}=\{\Omega^{\rm locked}_i\}$ is found, we then
fix it as the natural frequency distribution and change the
initial conditions of $\{\{\theta_i(0)\},\{\omega_i(0)\}\}$. With
the set $\{\{\Omega^{\rm locked}_i\}, \{\theta^{(1)}_i(0)\},
\{\omega^{(1)}_i(0)\}\}$, $\{\{\Omega^{\rm locked}_i\},
\{\theta^{(2)}_i(0)\}, \{\omega^{(2)}_i(0)\}\}$, and so on. The
gathered 100 phase-locked states of final phases,
$\{{\theta^{(1)f}_i}\}$, $\{{\theta^{(2)f}_i}\}$,...,
$\{{\theta^{(100)f}_i}\}$, are plotted as $z_i\equiv
e^{i\theta^f_i+\Delta\theta}$ on the complex plane as shown in
Fig.~\ref{fig:frameAB_last_phase} where only three cases are
taken. Figure~\ref{fig:frameAB_last_phase} shows that different
initial conditions reach indeed the same phase-locked state with
the relative phase-shift, $(\Delta\theta)^{\infty}$, between one
and another, which are tested in Framework A and Framework B for
two different distributions, respectively.

The relative phase-shift is the amount of a rotated angle, which
is needed to collapse numerical data of several phase-locked
states with different initial conditions but the same natural
frequency distribution. Figure \ref{fig:frameAB_shift} shows that
all 100 sets of final phases coincide with one another. The sum of
errors, $\sum_{i=1}^{N}{|\theta^{(j)f}_i-\theta^{(k)f}_i|}$, is
less than $10^{-10}$, where $i$ is an oscillator index and $j, k$
are sample indices. In Fig.~\ref{fig:frameAB_shift}, numerically
estimated values of $(\Delta\theta)^{\infty}$ are compared with
analytically estimated values (shift prediction) using initial
conditions, based on Theorem~\ref{OrbitalStability}, which implies
that the analytic estimate is correct.

Finally, in
Figs.~\ref{fig:Noframe_E_vs_t},~\ref{fig:Noframe_last_phase}, and
~\ref{fig:Noframe_shift}, we numerically show that arbitrary
initial conditions (neither Framework A nor Framework B) also
converge to an unique and stable phase-locked state as long as
keeping the same natural frequency distribution. This implies that
the restrictions of Framework A and Framework B are not necessary
for the system to converge to a phase-locked state for a given
natural frequency distribution. All the conditions, parameters,
values used in our simulation tests are shown in
Table~\ref{table:All-Conditions} and
Table~\ref{table:Three-Cases}.

\section{Conclusion}
\setcounter{equation}{0}

In summary, we have presented a simple proof for the nonlinear
stability of some classes of phase-locked states in terms of the
modified Kuramoto model with finite inertia in the context of
$\ell^{\infty}$-norm. Phase-locked states of the modified Kuramoto
model with $\Omega_c=0$ correspond to the equilibrium solutions of
the system \eqref{main}:
\begin{equation} \label{phase-equation}
 \Omega_i + \frac{K}{N} \sum_{i=1}^N \sin(\theta_j - \theta_i) = 0,
\qquad {\dot \theta}_i = 0.
\end{equation}

The solvability of the above nonlinear equation is not clear at
all. In the earlier work~\cite{C-H-Y}, the existence of
phase-locked states are established via a {\it time-asymptotic
approach}, i.e., instead of solving the above nonlinear system
\eqref{phase-equation} directly, the time-dependent system
\eqref{main} was solved from some admissible initial
configurations. Therefore, in the time-asymptotic limit, the
desired nonlinear system \eqref{phase-equation} can be solved.
This time-asymptotic approach can reveal the fine structure of the
phase-locked states (see \cite{C-H-J-K} for the original Kuramoto
model). Moreover, two frameworks were presented for the
well-selected parameters and initial configurations to guarantee
the validity of this time-asymptotic approach.

In this paper, we have adopted the same framework ideas employed
in the aforementioned work, where the existence of phase-locked
states were only investigated, and showed that the phase-locked
states whose existence is guaranteed by \cite{C-H-Y} are orbitally
$\ell^{\infty}$-stable in the sense that its small perturbation
leads to the relative phase-shift from the original phase-locked
state. Based on this fact, we claim that the phase-locked state
has in some sense a robust structure. This implies why we can
often observe a typical phase-locked states in biological systems.
As can be seen in numerical simulation tests (see
Fig.~\ref{fig:frameAB_last_phase}), Framework A and Framework B in
Sec.~\ref{proof} do not generate all the possible phase-locked
states and our stability theory do not cover the phase-locked
states whose phase-diameter is larger than $\frac{\pi}{2}$. At
present, we cannot provide a complete classification for the
stability of phase-locked states to the system~\eqref{main}, such
as the clear answers of the following questions: For a given $m,
K$ and $\{\Omega_i \}$, how many phase-locked states exist? Are
all phase-locked states orbitally stable? Of course, it is
impossible to find unstable phase-locked states in our numerical
investigation because intrinsic numerical errors make unstable
phase-locked states invisible. Thus, we leave these intriguing
issues for future works to be investigated~\cite{CCHHK} as well as
the nature of phase transitions in the modified Kuramoto
model~\cite{CHHK}.

Finally, based on our extensive numerical simulation tests, we
also note that the restriction of the phase-diameter in
phase-locked states is no longer necessary to support the orbital
stability, $D(\theta^{e})\le \frac{\pi}{2}$, which is related to
the framework setup. In other words, we numerically observe that
the phase-locked state with $D(\theta^{e})> \frac{\pi}{2}$
satisfies its uniqueness and orbital stability as well.

\section*{Acknowledgements}

The work was supported by the National Research Foundation of
Korea (NRF) grant funded by the Korean Government (MEST) (No.
2011-0011550) (M.H.), partially supported by the NRF grant MEST
(No. 2011-0015388) (S.Y.H.), and partially supported by the NRF
grant awarded through the Acceleration Research Program (No.
2010-0015066) and the NAP of KRCF (C.C.).


\end{document}